\documentclass[12pt]{article}

\usepackage{graphicx}
\usepackage{color}

\input epsf
\def\beq{\begin{eqnarray}}
\def\eeq{\end{eqnarray}}

\def\L*{{\cal L}_*}
\def\lsim{\mathrel{\rlap{\lower3pt\hbox{\hskip0pt$\sim$}}
     \raise1pt\hbox{$<$}}}         
\def\gsim{\mathrel{\rlap{\lower4pt\hbox{\hskip1pt$\sim$}}
     \raise1pt\hbox{$>$}}}         
\def\t{\tilde}

\usepackage{amsmath}
\usepackage{amsfonts}

\begin{document}

\begin{titlepage}

\begin{flushright}
{NYU-TH-07/12/01}
\end{flushright}
\vskip 0.9cm

\centerline{\Large \bf Mass Screening in Modified Gravity}

\vskip 0.7cm
\centerline{\large Gregory Gabadadze$^{a,}$\footnote{E-mail: gg32@nyu.edu} 
and Alberto Iglesias$^{b,}$\footnote{E-mail: iglesias@physics.ucdavis.edu}}
\vskip 0.3cm
\centerline{$^a$\em Center for Cosmology and Particle Physics}
\centerline{\em Department of Physics, New York University, New York, 
NY, 10003, USA}
\bigskip
\centerline{$^b$\em Department of Physics, University of California, Davis,
CA 95616}

\vskip 1.9cm

\begin{abstract}

Models of modified gravity introduce extra degrees of freedom, which 
for consistency with the data, should be suppressed at observable 
scales. In the models that share properties of massive gravity such 
a suppression  is due to nonlinear interactions:
An isolated massive astrophysical object creates a halo of a nonzero 
curvature around  it, shielding  its vicinity from the influence of the 
extra degrees of freedom.  We emphasize that the very same  halo 
leads to a screening  of the gravitational mass of the object, 
as seen by an observer beyond the halo. We discuss the case when  
the screening could be very significant and may rule out, or render the models 
observationally interesting.

\end{abstract}

\vspace{3cm}

\end{titlepage}

\newpage

\section{Introduction and summary}

One of the most puzzling discoveries of our times is the fact that the 
present-day expansion of the Universe is accelerating \cite {Acc}. Such an 
acceleration can be attributed to the existence of   
``dark energy''- a substance with a negative enough pressure
-  that is present in the Universe and, hence, in the 
rhs of the Einstein equation:
\beq
G_{\mu\nu} = 8\pi G_N (T^{\rm matter}_{\mu\nu}+ 
T^{\rm dark~ energy}_{\mu\nu})\,,
\label{Einstein}
\eeq
where $G_{\mu\nu}$ stands for the Einstein tensor
of the 4D space-time with metric $g_{\mu\nu}(x)$, and 
$T^{\rm matter}_{\mu\nu} $ and $T^{\rm dark~ energy}_{\mu\nu}$
denote the stress-tensors for visible and dark matter, 
and dark energy, respectively.

On the other hand, one can consider a different  
logical possibility: that the accelerated expansion 
is due to modified General Relativity (GR)\footnote{The latter 
approach is motivated by the ``old 
cosmological constant problem'' (for a review of which, 
see, e.g., \cite {Weinberg}, and in the context of modified gravity 
see, e.g., \cite {GGrev}). We will not be discussing 
this problem in the present work.}. In this case, we would have the modified 
Einstein equations of the form:
\beq
G_{\mu\nu} - {\cal K}_{\mu\nu}(g,m_c) = 
8\pi G_NT^{\rm matter}_{\mu\nu}\,,
\label{MGR}
\eeq
where  ${\cal K}_{\mu\nu}(g,m_c)$ denotes a tensor that could depend 
on a metric $g$, its derivatives, as well as on other fields not present in 
GR. Moreover, ${\cal K}$ depends on a dimensionful constant
$m_c\sim H_0\sim 10^{-42}~GeV$, that sets the distance/time  
scale $r_c\equiv m^{-1}_c$ at which the Newtonian potential significantly 
deviates from the conventional one.  For instance, in the DGP model 
\cite {DGP} ${\cal K}_{\mu\nu}$ is related to the extrinsic curvature 
tensor that  gives rise to a self-accelerated solution 
\cite {Cedric,DDG} (see, the comments on viability of this 
solution at the end of this section, and Ref. \cite {GGCargese} 
for a recent review). 

Even though the difference between  (\ref {MGR}) and (\ref {Einstein}) 
might seem just conventional at a first sight,
in reality, however, it could be observationally  significant.  
For instance, it is possible to choose the rhs of (\ref{Einstein})
so that it gives rise to a  background evolution 
obtained from (\ref{MGR}), nevertheless, 
perturbations on these backgrounds would be different in 
(\ref {MGR}) and (\ref {Einstein}), see, e.g., 
\cite {Lue2}-\cite{KK}.  

In what follows we will focus on the issue of weather the two approaches, 
(\ref {MGR}) and (\ref {Einstein}), could be differentiated 
by properties of a Schwarzschild-like  solution for a static 
spherically symmetric source. 
 
A key feature of any theory of modified gravity of the form (\ref{MGR}) 
is that, unlike GR, it allows for the possibility of having 
non-vanishing curvature outside of a source. 
This can be  easily understood by taking the trace of (\ref{MGR}) 
in a region outside of sources (where $T_{\mu\nu}=0$),
that gives:
\beq
-R={\cal K}~,
\eeq
where ${\cal K}={\cal K}_\mu^\mu$  needs not be zero. 
This fact affects in particular the notion of mass and 
leads to the interesting phenomenon of {\em screening}.
For example, when defining the Komar mass (which coincides with the ADM mass 
in stationary asymptotically flat space-time) one starts with an expression in 
terms of a volume integral of a projection of the  
Ricci tensor (see, e.g., \cite {Wald}) and then by use of 
Einstein's equations the Ricci tensor is replaced by 
$T_{\mu\nu}-g_{\mu\nu}T/2$. In a modified gravity theory of the form 
(\ref{MGR}) an extra term containing ${\cal K_{\mu\nu}}$ and its trace 
is generated in the replacement of the Ricci tensor. Thus, the definition of 
the Komar mass contains an extra piece, referred to as a mass deficit below, 
which for a compact static source of spherical symmetry is given by 
\beq
\Delta M\propto M_{Pl}^2\int dr^3 
\left ( {\cal K}_{00}-{1\over 2}g_{00}{\cal K}\right)~.
\label{deltaM}
\eeq
How significant is the mass deficit? The answer to this question would 
depend on a concrete  model at hand. However, we would like to argue that
in models which share properties of Lorentz invariant  
``massive gravity'', the mass deficit could be of the order of the 
mass itself.  This has something to do with the fact that 
such models exhibit the so called  strongly coupled behavior 
\cite {Arkady,DDGV}  (see also \cite{AGS,Rat1,Rubakov,Rat2}, 
and section 2 below for a summary), in spite of the fact that 
gravitational fields is weak everywhere 
\cite {DDGV,Gruzinov,Tanaka,NemanjaSh}.
 
\begin{figure}[h]
	\centering
	\includegraphics[width=8cm]{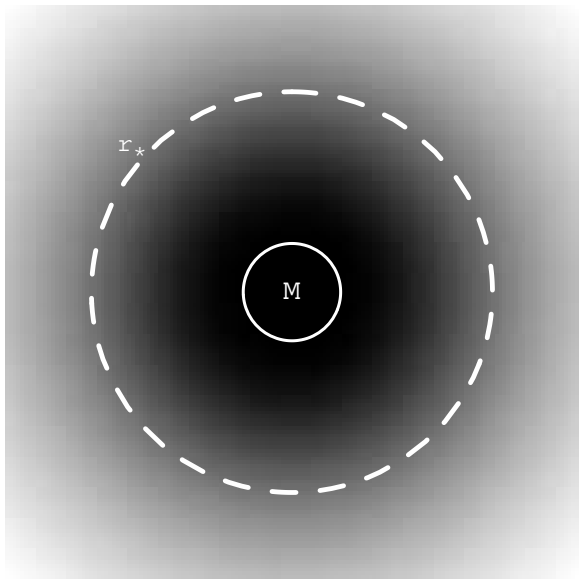}
	\caption{\small Curvature extends outside a source 
to a distance $r_*$.}
	\label{fig1}
\end{figure}

One way to interpret this property for a static spherically symmetric 
source  is to observe \cite {GI} that the source gives rise to the 
curvature that extends up to a macroscopic distance $r_*$, in the way 
depicted in figure \ref{fig1}. This curvature contributes to 
the integral (\ref {deltaM}). 

In a concrete case of the non-perturbative Schwarzschild solution of 
the DGP model \cite {GI}, we will show that 
this integral is saturated around a  distance $r_*$ 
($\sim (M /m_c^2M_{Pl}^2)^{1/3}$ - the topic of the next section )
where the value of ${\cal K}$ is typically of order $m_c^2$.
Then, the result of (\ref {deltaM}) is $$\Delta M\sim M~,$$
{\it i.e.}, the contribution to the mass from the modification of 
gravity term is of the same order as the mass itself!

In the following sections we will make concrete the statements pointed out 
above in the DGP model of modified gravity. They include the perturbative 
arguments leading to the derivation of the scale $r_*$ and details 
on the exact solutions available for Schwarzschild-like sources and domain 
walls. While the former are based on an {\it ansatz}, legitimate concerns 
about the bulk boundary conditions of which were raised in \cite {KKBoundary},
nevertheless, the fact that the {\it ansatz}  recovers  very  precisely 
the 4D GR results at short distances  and smoothly interpolates to the 
5D regime  (that no other solution is known to do), suggest that it 
may  be capturing right physics. Moreover, the above 
properties were  subsequently found to be true for  the case of a 
domain wall for which an exact solution was obtained \cite {DGPR}.

Although the existence of the mass deficit (\ref {deltaM}) could be 
interesting observationally, it may  lead in certain cases  to 
problems with the theory. Indeed, because of the terms  (\ref {deltaM}) in 
the expression  for a  gravitational mass the proof of 
the positive energy theorem 
\cite {Yau,Witten}  is not directly applicable. Hence, in general, 
there could exist  negative ``mass'' solutions  \cite {Cedric,Rat1} even 
for matter stress-tensors that satisfy conventional positive energy 
conditions.  One example of this 
is the self-accelerated solution \cite {Cedric,DDGV}. Small perturbations 
about this solution in a linearized theory and 
with non-conformal sources exhibit ghost-like states 
\cite {Rat1,Rat2,Gorbunov,Kaloper,KKTanaka}, 
however, there exist serious arguments that the  
perturbative results  cannot be trusted in the full non-linear 
theory \cite {DGI} (see also \cite {Dvali}). 
Nevertheless, some semi-exact \cite {GI} and 
exact \cite {DGPR,Kaloper,KaloperMyers}  non-perturbative solutions 
on the self-accelerated branch exhibit ``negative mass''.
This suggest that the self-accelerated branch should be unstable,
however, it is not clear what is the time of its instability.
An explicit calculation on decay of the selfaccelerated branch into 
the conventional one shows that such a decay does not take place,
at lest in a quasi-classical approximation \cite {Bounce}.
This question is still open and we will not be discussing it further 
in the present work.

If the mass screening is substantial, then,
at scales beyond $r_*$ gravity would be modified significantly.
However, the value of the scale $r_*$ for the entire 
observable Universe is $H_0^{-1}\simeq 10^{28}$ cm. 
Therefore, on average, the beyond-$r_*$-effects will 
be hard to detect. There  may be exceptions  for isolated
clusters of galaxies separations between which are 
greater than their own  $r_*$ scales, and any other $r_*$ 
scales in their vicinity\footnote{We thank 
Lam Hui and Roman Scoccimarro for discussions 
on these issues.}.  For precise calculations of the beyond 
$r_*$ physics, however, new averaging technique and 
non-perturbative calculational methods would be needed.

\section{The $r_*-$scale (Vainshtein scale)}

In what follows we will concentrate on the concrete example provided by 
the DGP model \cite {DGP} in which all interactions except gravity are 
confined to a 4D brane embedded in an infinite volume 5D empty space where 
gravity propagates.

The modification of gravity in this model is given in terms of the extrinsic 
curvature $K_{\mu\nu}$ of the brane and reads:
\beq\label{mod}
{\cal K_{\mu\nu}}=m_c(K_{\mu\nu}-g_{\mu\nu}K)~.
\eeq  
The 5D space has coordinates $(x^\mu,y)$ with the brane at the surface 
$y=0$ and the 4D metric in (\ref{mod}) is $g_{\mu\nu}(x^\mu,y=0)$.
 
The macroscopic distance $r_*$ in the DGP model can be derived by considering 
the linearized analysis of the theory \cite {DGP}.
In particular we will focus on 
the one-graviton exchange amplitude between two sources whose gauge 
independent expression reads as follows: 
\beq
{\cal A}_{\rm 1-graviton}(p,\,y)\,=
\,{T^2_{1/3} \over p^2\,+\,m_c\,p}{\rm exp} (-p |y|)\,,
\label{A13}
\eeq
where 
\beq
T^2_{1/3} \,\equiv\, 8\,\pi\,G_N \left ( 
T^2_{\mu\nu}\,-\,{1\over 3}\,T\cdot T \right )~,
\label{T1third}
\eeq
and $p^2$ is the square of the Euclidean brane 4-momentum. 
The pole at $p^2=0$ has zero residue, and the 
second pole in  (\ref {A13})  is on a non-physical Riemann sheet. The former 
implies the absence of a massless graviton in the exchange while the latter   
describes the propagation of a metastable state
with lifetime $\sim m^{-1}_c$, which decays into a 
continuum of KK modes.

The striking feature of (\ref{T1third}) is that in the  
$m_c\to 0$ limit (in which the modification of gravity should disappear)
the numerator does not reduce to the analogous expression in GR:
\beq
8\,\pi\,G_N \left ( 
T^2_{\mu\nu}\,-\,{1\over 2}\,T\cdot T \right )\,.
\label{T1half}
\eeq
This fact, which could be used to exclude (\ref{T1third}) by observations,
is known as the van Dam-Veltman-Zakharov discontinuity (vDVZ) \cite {vDVZ}.
The difference between  (\ref {T1third}) and 
(\ref {T1half}) is due to the fact that a 5D graviton (or a 
massive graviton for that matter) propagates 5 on-shell degrees of 
freedom (helicity-2, helicity-1, and helicity-0), while the GR 
graviton propagates only 2 on-shell degrees of freedom 
(helicity-2 state). And while the helicity-1 state of the 5D graviton  
does not contribute to 
(\ref {A13}) at the linearized level because of the contraction with 
conserved sources, the helicity-0 state couples to the trace of the 
energy-momentum tensor and gives a non-vanishing contribution 
(when $T\not =0$).

It has been argued in \cite {Arkady,DDGV} that the continuity in the 
$m_c\to 0$ limit would be restored if nonlinear effects were taken into 
account. The relevance of these terms can be understood in the following 
terms: the longitudinal  part of the graviton 
propagator in DGP contains terms proportional to 
$p_\mu p_\nu/m_c p$.
This term does not contribute to the amplitude (\ref{A13}) 
because of conservation of the stress-tensor,
but it does contribute already in the first  nonlinear 
correction (since the stress-tensor is only covariantly conserved
in the non-linear theory). And due to the singular behavior of these 
terms in the  
$m_c\to 0$ limit, perturbation 
theory  breaks down  prematurely.
However, this breakdown is  an artifact of an ill-defined 
perturbative expansion -- the  known  exact solutions of the 
model have no trace of breaking \cite {DDGV}.
The perturbative expansion in powers of $G_N$ 
gets ``contaminated'' by another dimensionful 
parameter $1/m_c$, and this leads to its breakdown.
As possible ways forward one could either
adopt a different type of expansion, e.g., an expansion 
in the small parameter $m_c$ \cite {DDGV,Gruzinov}, 
or try to find exact solutions
\footnote{It is also possible to modify the theory at the linearized 
level so that the conventional perturbative expansion is well-behaved 
\cite {GGweak,Siopsis},\cite {Massimo},\cite{Smolyakov}.}.
Both of these programs have been carried out to a certain  
extent and we will review in the next section the salient features of the 
latter.

As presented in (\ref{MGR}), the DGP model has one adjustable parameter, namely
$m_c$ which determines a scale that separates two different regimes of the 
theory. 
For distances much smaller than $m_c^{-1}$ one would expect the solutions to 
be well approximated by GR and the modifications to appear at larger distances.
This is indeed the case for distributions of matter and  radiation which are 
homogeneous and  isotropic at scales $ \gsim r_c$ \cite {Cedric,DDG,DDGV}.
However, more compact sources exhibit different properties. For example,
a compact static source 
of the mass $M$ 
and radius $r_0$, such that $r_M < r_0  \ll r_c$ 
($r_M\equiv 2G_NM$  is the  Schwarzschild  radius) a new scale, combination of
 $r_c$ and $r_M$, emerges 
(the so-called  Vainshtein scale\footnote{A similar, but not 
exactly the same scale was discovered by Vainshtein in massive 
gravity \cite {Arkady}, hence the name.}) \cite {DDGV}: 
\beq
r_*\equiv (r_M r_c^2)^{1/3}\,.
\label{r*}
\eeq
Below this scale the predictions of the theory are in 
good agreement with the GR results and above it they deviate considerably. 
These type of sources will be discussed in more detail in the next section 
together with other ones with higher simpler symmetry: domain walls.


\section{Concrete examples}

In this section we will focus on the mass screening phenomenon describing how 
it arises in the cases of the Schwarzschild-like non-perturbative solution 
(NPS) of \cite{GI,GIlunar} and the exact domain wall (DW) solutions of 
\cite {DGPR}.

\subsection{Schwarzschild solution}
{}The NPS solution studied in \cite{GI, GIlunar} 
is found by considering a static metric with spherical symmetry on the
brane and with ${\bf Z}_2$ symmetric line element:
\begin{equation}
ds^2=-{\rm e}^{-\lambda} dt^2+{\rm e}^\lambda dr^2+r^2d\Omega^2+2\gamma\ dr
dy+{\rm e}^\sigma dy^2~,
\label{interval}
\end{equation}    
where $\lambda,\ \gamma,\ \sigma$ are functions of the radial coordinate on 
the brane $r$ and the transverse direction coordinate $y$. The ${\bf Z}_2$ 
symmetry across the brane ($y=0$) implies that $\gamma$ is 
an odd function of $y$ while the rest are even. The choice 
$-g_{tt}=1/g_{rr}$ represents an {\it ansatz}, but notice that we have kept 
the off-diagonal term $g_{ry}=\gamma$. 

The brane is chosen 
to be straight in the above coordinate system but one could 
transform (\ref {interval}) to another one in which the metric 
is diagonal $ds^2 =-A(r,z)dt^2 + B(r,z)d\rho ^2 + C(r,z) d\Omega^2 +dz^2$. 
Here, $A\not =1/B$ in general, and the ansatz is reflected in 
the fact that in this system our brane will be bent or, in other words, in 
a particular nontrivial choice of the position of the brane $z(r)$.

This ansatz allows us to close the system of equations {\it on the brane}, 
and leads to the following solutions for the gravitational 
potential
$\phi$,
\beq
{\rm e}^{-\lambda} &=& 1+2\phi~,\\
\phi(r)&=&{3m_c^2\over 4r} \int {\rm d}r\ r^2 U(r)~. 
\label{P}
\eeq
The function $U(r)$ in (\ref{P}) is given implicitly by the solutions of the 
following two equations 
(giving rise to a conventional  and self-accelerated branch respectively): 
\begin{eqnarray}
(k_1 r)^8&=&-{(1+3U+f)\over U^2(3+3U+\sqrt{3}f)^{2\sqrt{3}}(-5-3U+f)}~,
\label{sol1}\\
(k_2 r)^8&=&-{(-5-3U+f)(-3-3U-\sqrt{3}f)^{2\sqrt{3}}\over (U+2)^2 
(1+3U+f)}~,\label{sol2}
\end{eqnarray}
where $f=\sqrt{1+6U+3U^2}$ and $k$ is an integration constant. 

The off-diagonal and $yy$ metric components are 
determined from 
\beq
&&{4r^2(r\phi)_r\over (r\phi)_{rr}}=
{\left(r^4\gamma {\rm e}^{-\lambda}\right)_r\over 
\left(r\gamma {\rm e}^{-\lambda}\right)_r}~,\\
&&{\rm e}^\sigma =-m_c^2\left[{\left(r^4\gamma{\rm e}^{-\lambda}\right)_r\over
4r^2(r\phi)_r}\right]^2+{\rm e}^{-\lambda}\gamma^2~,
\eeq
and the profile of the warp factors (``y-derivatives'') can be computed on the 
brane as well.

There are two integration constants, $k$ and the one produced in the 
integration (\ref{P}), which are determined by imposing appropriate boundary 
conditions near the source ($r\ll r_*$) and at large distances. 
For the first condition we impose the 4D behavior of the potential near the 
source: $\phi=G_NM/r$, while for the second one we require that the 
coefficient of the possible $1/r$ term be zero, {\it i.e.}, to be left with 5D 
behavior at large distances, 
namely, $\lambda\sim \t r_M^2/r^2$  in the conventional branch and 
$\lambda \sim m_c^2r^2+\t r_M^2/r^2$ in the self-accelerated branch.

\subsubsection{Conventional branch}

{}The conventional branch is obtained from the solution of (\ref{sol1}).
As shown in \cite{GI, GIlunar} the boundary conditions discussed above  
 determine the  
asymptotic behavior of the solution. At 
short distances, $r\ll r_*$ ($U\to +\infty$), we get
\beq\label{short}
\phi=-{G_NM\over r}+{1\over 2}\alpha_1 m_c^2r^2 
\left({r_*\over r}\right)^{2(\sqrt{3}-1)}+\dots~,
\eeq
where $\alpha_1\approx 0.84$ and the coefficient of the $1/r$ term was chosen 
to be $-G_NM$ by fixing the constant of integration in (\ref{P}). 

The other integration constant $k_1$ is chosen such that at
large distances, $r\gg r_*$ ($U\to 0^+$), we obtain an expansion with no $1/r$
term:
\beq
\phi =-{\t r_{M_1}^2\over2 r^2}+\dots~,
\eeq
which fixes the value of $k_1$ in terms of $r_*$ 
($(r_*k_1)^3\approx 0.21$) and also implies 
\beq\label{trm}
\t r_{M_1}^2\approx 0.56\ r_M r_*~.
\eeq

The relation (\ref{trm}) should be contrasted with the naive expectation 
from a linearized analysis: in the 5D regime one would have expected to have 
$\t r_{M_1}\sim r_M r_c$, however, we get a much smaller value, reduced by a 
factor $r_*/r_c\equiv (r_M/r_c)^{1/3}$.

\begin{figure}[h]
	\centering
	\includegraphics[width=8cm]{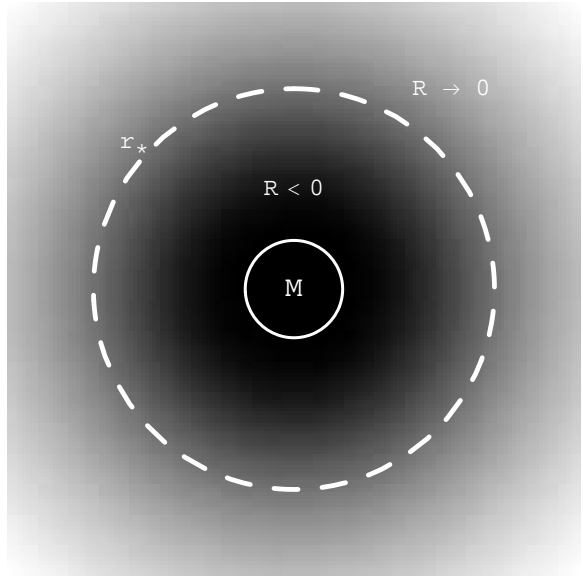}
	\caption{Conventional branch}
	\label{fig2}
\end{figure}

Therefore, as we see, a short distance observer at $r_M \ll r\ll r_*$ 
would measure the gravitational mass $M$ with a small corrections
to Newton's potential, while the large distance observer
at $r\gg r_*$  would measure an effective gravitational mass
$ \sim M(r_M/r_c)^{1/3}$ \cite {GI}. The latter 
includes the effects of the 4D curvature which is significant up to $r_*$ as 
depicted in figure \ref{fig2}. The 5D mass is partially screened at large 
distances.

Another point worth emphasizing is that a perturbative expansion 
suggests that for $r_*
\ll r \ll r_c$ the metric should have an approximately
four-dimensional, $1/r$, scalar-tensor-gravity type form
\cite{DGP,DDGV}. However, the NPS above 
exhibits a different behavior: beyond $r_*$ the
metric turns into the one produced by a five dimensional source.
We interpret this as a complete screening of the 4D mass of the source 
by the halo of non-zero curvature.

\begin{figure}[h]
	\centering
	\includegraphics[width=8cm]{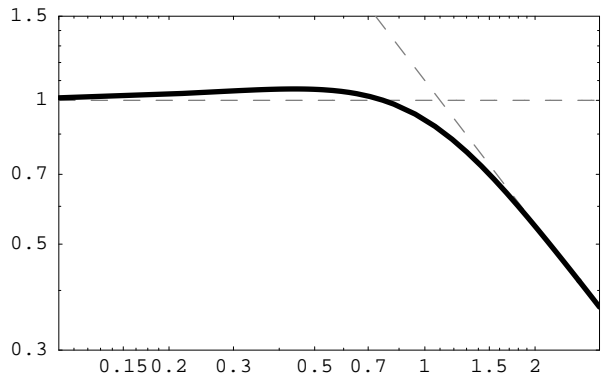}
	\caption{\small Log-Log plot of $M_{eff}/M$ vs. $r/r_*$. The oblique 
dashed line shows a $1/r$ fall-off.}
	\label{figmass}
\end{figure}

The screening of the 4D mass can be made explicit by taking into account the 
expression for the 4D Komar mass as a function of $r$. In the ansatz
used here gives this gives an effective mass
\beq\label{meff}
M_{eff}\sim M_{Pl}^2\left(-r\phi+r(r\phi)_r\right)~.
\eeq 
The first term the rhs of (\ref{meff}) is a smooth decreasing function 
of $r$ and gives a contribution that is $\sim M$ (the original mass of 
the source)
up to $r_*$ and rapidly falls off like $1/r$ beyond that point.
The second term, on the other hand, depends on the gradient of $\phi$ and is 
peaked around $r_*$. The combined effect is seen in figure \ref{figmass}:
the effective mass increases from its 4D value $M$ near the source up to 
$r\sim r_*$ and then falls to zero abruptly. 

The short distance mass increase
can also be deduced form the approximate form of the potential (\ref{short}) 
since the second term provides an additional attraction toward the source.

\subsubsection{Self-accelerated branch}

{}The solution on the self-accelerated branch is obtained from (\ref{sol2}).
The relation between $k_1$ and $r_*$ is obtained, as in the conventional case,
by imposing boundary conditions together with the following 
asymptotic behavior. At large distances, $r\gg r_*$ ($U\to -2^-$), we 
derive 
\beq
\phi={\t r_{M_2}^2\over2 r^2}-{1\over 2}m_c^2r^2+\dots~,
\eeq
where,
\beq
\t r_{M_2}^2\approx 0.45\ r_M r_*~,
\eeq
, {\it i.e.}, 5D mass screening. At short 
distances, $r\ll r_*$ ($U\to -\infty$), we get
\beq
\phi=-{G_NM\over r}+{1\over2}\alpha_2 m_c^2r^2 
\left({r_*\over r}\right)^{2(\sqrt{3}-1)}+\dots~,
\eeq
where $\alpha_2=-\alpha_1 \approx -0.84$ is, in absolute value, the same 
constant appearing in the conventional 
branch short distance expansion (\ref{short}). Note, however, that the sign of 
the correction to the $4D$ behavior is opposite in the two branches.

\begin{figure}[h]
	\centering
	\includegraphics[width=8cm]{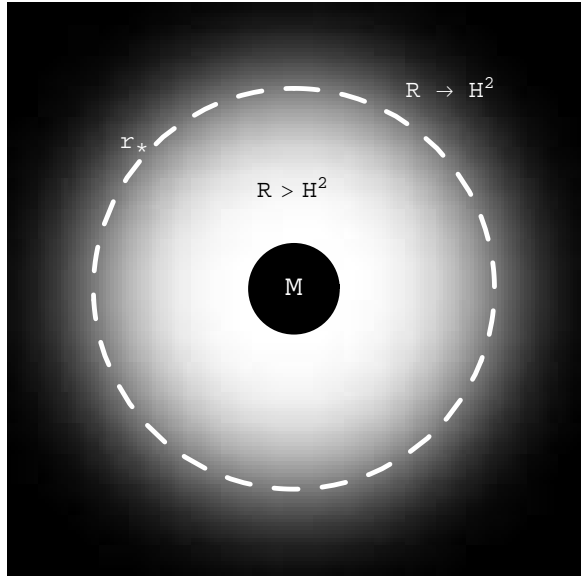}
	\caption{Self-accelerated branch}
	\label{fig3}
\end{figure}

At intermediate  distances, $r_* \ll  r \ll r_c$,  
the potential contains a $5D$ gravitational term that 
is {\em repulsive}, ${\tilde r}^2_M/r^2$. 
This looks like a  5D negative mass. However, this is not an 
asymptotic value of the mass since one can 
only cover the solution in the above  coordinate system till 
$r\sim r_c$ where the dS like horizon is encountered. 
Moreover, in the intermediate regime
$r_* \ll  r \ll r_c$, the  de Sitter term $m_c^2 r^2$ 
in the potential always dominates over the 
${\tilde r}^2_M/r^2$ term suggesting that the effects 
due to the Schwarzschild source are strongly suppressed.
The picture that explains screening in this branch is depicted 
in figure \ref{fig3}.

\subsection{Domain Walls}

The second example that illustrates the screening phenomenon in the DGP
model is that of a brane DW source. The study of this type of 
sources was done in \cite{DGPR} on which we base the following 
discussion.  

The source considered is a Nambu-Goto DW in 4D localized at $z=0$ 
with stress-tensor
\beq
T_{\mu\nu}=\sigma \, \delta(z) \;{\rm diag}(1,-1,-1,0)~,
\label{dwt}
\eeq
where $\sigma$ is the tension of the wall and $z$ denotes the
coordinate transverse to its world-volume spanned by ($t,x,y$). 

The domain wall solution for such a source in GR \cite{Vilenkin,IS} displays 
3D de Sitter expansion in its world-volume at a rate  $H=2\pi G_N \sigma$.

In DGP, however, the situation is different. For tensions smaller 
than a critical value, $\sigma_c\equiv M_*^3=m_c M_{Pl}/2$, the wall has no 
gravitational effects. One way to understand this is by noticing that for 
these sources, the  modification of
gravity term (\ref{mod} ) precisely compensates the energy momentum tensor 
$T_{\mu\nu}$. Therefore,
the tension of the wall, as seen from the point of view of a 4D
observer, is screened entirely by gravitational effects encoded in
the extrinsic curvature. Not surprisingly, the domain wall world-volume
remains flat, and so does the metric on the brane.

Furthermore, this screening
takes place inside the core of the wall. Hence, the analog notion of the
$r_*$ scale for a domain wall (understood as where the self-shielding takes 
place) coincides with its thickness,
$$
r_*^{(DW)}=d_{core}~.
$$
This is to be compared to the Schwarschild-like case, where
the shielding also occurs, and $r_*$ extends outside the source.
The net result is the screening of the 4D tension/mass in both cases.

For supercritical tension branes the extrinsic curvature can no longer
balance the energy momentum tensor and the brane inflates. The transverse 
direction to the wall is compactified to a size $d$ and a zero mode graviton 
appears. Since this phenomenon also takes place in GR, it provides a 
means to contrast the 5D effects in DGP with those of a supercritical DW
in 5D, {\it i.e.}, by comparing the world-volume inflation rate in both cases.

The exact solutions of \cite{DGPR} give a  suppression factor for the 
inflation rate of the supercritical DW in DGP for both branches of solutions. 
In the conventional branch it is $d/2r_c$ while in the self-accelerated 
branch it is given by $-d/2r_c$. As argued above, we should identify the 
Vainshtein scale $r_*$ for these sources $d$ and therefore we find the same
parametric screening of the 5D tension as for the mass in the 
Schwarzschild-like solutions of the previous example.

\section{Discussions}

The above results may also be applicable to other models where 
the results of 4D gravity are recovered through strongly coupled 
behavior. The minimal model of brane induced gravity in greater that
five dimensions \cite {DG} contains ghosts \cite {RD}. However, its
extensions are ghost free  \cite {GS}, a small subset of which 
has a strong coupling regime \cite {GS,StrongSoft} (see, also 
\cite {NemanjaH}).  
A recent model of cascading brane induced gravity  is also ghost 
free \cite {Cascade}.  It would be interesting 
to understand the issue of presence/absence of the 
mass screening   in these models (see, also, \cite {Koy}).

Unfortunately, at present  there is no consistent 4D theory of  
Lorentz invariant massive gravity, as it suffers from nonlinear instabilities
\cite{Deser,GGruzinov,CedricRomb,Nicolis}.  However, the mass screening 
effect described above is based on rather universal principles, and 
it is reasonable to expect that the phenomenon  will 
also persists  if  a consistent model is found.

It would  be interesting to understand the issue of the mass screening
in the $f(R)$-type models  of models of modified gravity, 
see, e.g., \cite {Bean}, as well as in Lorentz  violating 
models  \cite {NimaL,RL,D,Tin,GGrisa}. Some of these models    
\cite {RL,D,Tin,GGrisa} avoid the strong coupling behavior, in which 
case the mass screening is not expected to be very significant.

 \subsection*{Acknowledgments}

GG would like to thank the organizers of the international 
workshop Peyresq-12 for warm hospitality and productive 
atmosphere. GG was supported by NASA grant NNGG05GH34G. AI was 
supported by DOE Grant DE-FG03-91ER40674.


\begin{thebibliography}{99}

\bibitem{Acc}  A.~G.~Riess {\it et al.}  [Supernova Search Team Collaboration],
  Astron.\ J.\  {\bf 116}, 1009 (1998)
  [arXiv:astro-ph/9805201]; \\
S.~Perlmutter {\it et al.}  [Supernova Cosmology Project Collaboration],
  Astrophys.\ J.\  {\bf 517}, 565 (1999)
  [arXiv:astro-ph/9812133].


\bibitem{Weinberg}
S.~Weinberg,
  Rev.\ Mod.\ Phys.\  {\bf 61}, 1 (1989).


\bibitem{GGrev}
G.~Gabadadze, 
  arXiv:hep-th/0408118; In Ian Kogan Memorial Volume, 
Shifman, M. (ed.) et al. World Scientific, 2004; vol.2, pp 1061-1130. 


\bibitem{DGP}
G.~Dvali, G.~Gabadadze and M.~Porrati,
Phys.\ Lett.\  {\bf B485}, 208 (2000)
[hep-th/0005016].

\bibitem{Cedric}
C.~Deffayet,
Phys.\ Lett.\ B {\bf 502}, 199 (2001)
[arXiv:hep-th/0010186].

\bibitem{DDG}
C.~Deffayet, G.~R.~Dvali and G.~Gabadadze,
Phys.\ Rev.\ D {\bf 65}, 044023 (2002)
[astro-ph/0105068].

\bibitem{GGCargese} G.~Gabadadze,
  Nucl.\ Phys.\ Proc.\ Suppl.\  {\bf 171}, 88 (2007)
  [arXiv:0705.1929 [hep-th]].


\bibitem{Lue2} A.~Lue, R.~Scoccimarro and G.~Starkman,
  Phys.\ Rev.\ D {\bf 69}, 044005 (2004)
  [arXiv:astro-ph/0307034].
  Phys.\ Rev.\ D {\bf 69}, 124015 (2004).


\bibitem{Dev} D.~Jain, A.~Dev and J.~S.~Alcaniz,
  Phys.\ Rev.\ D {\bf 66}, 083511 (2002)
  [arXiv:astro-ph/0206224].
J.~S.~Alcaniz, D.~Jain and A.~Dev,
  Phys.\ Rev.\ D {\bf 66}, 067301 (2002).


\bibitem{Spergel} M.~Ishak, A.~Upadhye and D.~N.~Spergel,
  Phys.\ Rev.\ D {\bf 74}, 043513 (2006)
  [arXiv:astro-ph/0507184].


\bibitem{Linder} E.~V.~Linder,
  Phys.\ Rev.\ D {\bf 72}, 043529 (2005)
  [arXiv:astro-ph/0507263].


\bibitem{Maa} R.~Maartens and E.~Majerotto,
  Phys.\ Rev.\ D {\bf 74}, 023004 (2006)
  [arXiv:astro-ph/0603353].


\bibitem{Maart} K.~Koyama and R.~Maartens,
  JCAP {\bf 0601}, 016 (2006)
  [arXiv:astro-ph/0511634].


\bibitem{Carroll} I.~Sawicki and S.~M.~Carroll,
  arXiv:astro-ph/0510364.



\bibitem{Hu} Y.~S.~Song, I.~Sawicki and W.~Hu,
  arXiv:astro-ph/0606286.

\bibitem{KK} A.~Cardoso, K.~Koyama, S.~S.~Seahra and F.~P.~Silva,
  arXiv:0711.2563 [astro-ph].





\bibitem{Wald}
  R.~M.~Wald,
{\it  Chicago, Usa: Univ. Pr. ( 1984) 491p}


\bibitem{Arkady} A.~I.~Vainshtein,
  %
  Phys.\ Lett.\ B {\bf 39} (1972) 393.

\bibitem{DDGV} C.~Deffayet, G.~R.~Dvali, G.~Gabadadze and A.~I.~Vainshtein,
  Phys.\ Rev.\ D {\bf 65}, 044026 (2002)
  [arXiv:hep-th/0106001].


\bibitem{AGS} N.~Arkani-Hamed, H.~Georgi and M.~D.~Schwartz,
  Annals Phys.\  {\bf 305}, 96 (2003)
  [arXiv:hep-th/0210184].

\bibitem{Rat1} M.~A.~Luty, M.~Porrati and R.~Rattazzi,
JHEP {\bf 0309}, 029 (2003)
[arXiv:hep-th/0303116].

\bibitem{Rubakov}V.~A.~Rubakov,
  arXiv:hep-th/0303125.

\bibitem{Rat2} A.~Nicolis and R.~Rattazzi,
  JHEP {\bf 0406}, 059 (2004)
  [arXiv:hep-th/0404159].


\bibitem{Gruzinov} A.~Gruzinov,
  New Astron.\  {\bf 10}, 311 (2005)
  [arXiv:astro-ph/0112246].

\bibitem{Tanaka} T.~Tanaka,
  Phys.\ Rev.\ D {\bf 69}, 024001 (2004)
  [arXiv:gr-qc/0305031].

\bibitem{NemanjaSh}  N.~Kaloper,
  Phys.\ Rev.\ Lett.\  {\bf 94}, 181601 (2005)
  [Erratum-ibid.\  {\bf 95}, 059901 (2005)]
  [arXiv:hep-th/0501028];
  Phys.\ Rev.\  D {\bf 71}, 086003 (2005)
  [Erratum-ibid.\  D {\bf 71}, 129905 (2005)]
  [arXiv:hep-th/0502035].

\bibitem{GI}
G.~Gabadadze and A.~Iglesias,
  Phys.\ Rev.\ D {\bf 72}, 084024 (2005)
  [arXiv:hep-th/0407049]; 

\bibitem{KKBoundary}K.~Koyama and F.~P.~Silva,
  Phys.\ Rev.\  D {\bf 75}, 084040 (2007)
  [arXiv:hep-th/0702169].

\bibitem{DGPR} G.~Dvali, G.~Gabadadze, O.~Pujolas and R.~Rahman,
  Phys.\ Rev.\  D {\bf 75}, 124013 (2007)
  [arXiv:hep-th/0612016].



\bibitem{Yau} R.~Schon and S.~T.~Yau,
  Commun.\ Math.\ Phys.\  {\bf 79}, 231 (1981).


\bibitem{Witten} E.~Witten,
  Commun.\ Math.\ Phys.\  {\bf 80}, 381 (1981).


\bibitem{Gorbunov} D.~Gorbunov, K.~Koyama and S.~Sibiryakov,
  Phys.\ Rev.\  D {\bf 73}, 044016 (2006)
  [arXiv:hep-th/0512097].

 
\bibitem{Kaloper}  C.~Charmousis, R.~Gregory, N.~Kaloper and A.~Padilla,
  JHEP {\bf 0610}, 066 (2006)
  [arXiv:hep-th/0604086].

\bibitem{KKTanaka} K.~Izumi, K.~Koyama and T.~Tanaka,
  JHEP {\bf 0704}, 053 (2007)
  [arXiv:hep-th/0610282].


\bibitem{DGI} C.~Deffayet, G.~Gabadadze and A.~Iglesias,
  JCAP {\bf 0608}, 012 (2006)
  [arXiv:hep-th/0607099].

\bibitem{Dvali} G.~Dvali,
  arXiv:hep-th/0402130.


\bibitem{KaloperMyers}
R.~Gregory, N.~Kaloper, R.~C.~Myers and A.~Padilla,
  JHEP {\bf 0710}, 069 (2007)
  [arXiv:0707.2666 [hep-th]].


\bibitem{Bounce} K.~Izumi, K.~Koyama, O.~Pujolas and T.~Tanaka,
  Phys.\ Rev.\  D {\bf 76}, 104041 (2007)
  [arXiv:0706.1980 [hep-th]].


\bibitem{vDVZ} H.~van Dam and M.~J.~G.~Veltman,
  %
  Nucl.\ Phys.\ B {\bf 22}, 397 (1970); \\
V.~I.~Zakharov,
  %
  JETP Lett.\  {\bf 12} (1970) 312
  [Pisma Zh.\ Eksp.\ Teor.\ Fiz.\  {\bf 12} (1970) 447].


\bibitem{DGZ} G.~Dvali, A.~Gruzinov and M.~Zaldarriaga,
Phys.\ Rev.\ D {\bf 68}, 024012 (2003)
[arXiv:hep-ph/0212069].



\bibitem{Lue1} 
A.~Lue and G.~Starkman,
Phys.\ Rev.\ D {\bf 67}, 064002 (2003)
[arXiv:astro-ph/0212083].



\bibitem{Iorio1}  L.~Iorio,
  Class.\ Quant.\ Grav.\  {\bf 22}, 5271 (2005)
  [arXiv:gr-qc/0504053]; 
  JCAP {\bf 0509}, 006 (2005)
  [arXiv:gr-qc/0508047];
  JCAP {\bf 0601}, 008 (2006).


\bibitem{GIlunar}
G.~Gabadadze and A.~Iglesias,
  Phys.\ Lett.\ B {\bf 632}, 617 (2006)
  [arXiv:hep-th/0508201].

\bibitem{GIDec}
G.~Gabadadze and A.~Iglesias,
Phys.\ Lett.\ B {\bf 639}, 88 (2006) [arXiv:hep-th/0603199]. 



\bibitem{GGweak}
G.~Gabadadze,
  Phys.\ Rev.\  D {\bf 70}, 064005 (2004)
  [arXiv:hep-th/0403161].

\bibitem{Siopsis}
C.~Middleton and G.~Siopsis,
  Phys.\ Lett.\  B {\bf 613}, 189 (2005)
  [arXiv:hep-th/0502020].

\bibitem{Massimo}
M.~Porrati and J.~W.~Rombouts,
  Phys.\ Rev.\ D {\bf 69}, 122003 (2004)
  [arXiv:hep-th/0401211].

\bibitem{Smolyakov}
M.~N.~Smolyakov,
  Phys.\ Rev.\  D {\bf 72}, 084010 (2005)
  [arXiv:hep-th/0506020].




\bibitem{Vilenkin} A.~Vilenkin,
  Phys.\ Lett.\ B {\bf 133} (1983) 177.


\bibitem{IS} J.~Ipser and P.~Sikivie,
  Phys.\ Rev.\ D {\bf 30}, 712 (1984).





\bibitem{DG} G.~R.~Dvali and G.~Gabadadze,
  Phys.\ Rev.\ D {\bf 63}, 065007 (2001)
  [arXiv:hep-th/0008054].


\bibitem{RD} S.~L.~Dubovsky and V.~A.~Rubakov,
  Phys.\ Rev.\  D {\bf 67}, 104014 (2003)
  [arXiv:hep-th/0212222].


\bibitem{GS} G.~Gabadadze and M.~Shifman,
  Phys.\ Rev.\ D {\bf 69}, 124032 (2004)
  [arXiv:hep-th/0312289].

\bibitem{StrongSoft} Work in progress.

\bibitem{NemanjaH} N.~Kaloper and D.~Kiley,
JHEP {\bf 0705} (2007) 045
[arXiv:hep-th/0703190].
N.~Kaloper,
  arXiv:0711.3210 [hep-th].


\bibitem{Cascade}C.~de Rham, G.~Dvali, S.~Hofmann, J.~Khoury, 
O.~Pujolas, M.~Redi and A.~J.~Tolley,
  arXiv:0711.2072 [hep-th].

\bibitem{Koy} O.~Corradini, K.~Koyama and G.~Tasinato,
  arXiv:0712.0385 [hep-th].

\bibitem{Deser} D.~G.~Boulware and S.~Deser,
  Phys.\ Rev.\  D {\bf 6}, 3368 (1972).

\bibitem{GGruzinov} G.~Gabadadze and A.~Gruzinov,
  Phys.\ Rev.\  D {\bf 72}, 124007 (2005)
  [arXiv:hep-th/0312074].

\bibitem{CedricRomb} C.~Deffayet and J.~W.~Rombouts,
  Phys.\ Rev.\  D {\bf 72}, 044003 (2005)
  [arXiv:gr-qc/0505134].

\bibitem{Nicolis} P.~Creminelli, A.~Nicolis, M.~Papucci and E.~Trincherini,
  JHEP {\bf 0509}, 003 (2005)
  [arXiv:hep-th/0505147].

\bibitem{Bean} R.~Bean, D.~Bernat, L.~Pogosian, A.~Silvestri and M.~Trodden,
  Phys.\ Rev.\  D {\bf 75}, 064020 (2007)
  [arXiv:astro-ph/0611321].

\bibitem{NimaL}
N.~Arkani-Hamed, H.~C.~Cheng, M.~A.~Luty and S.~Mukohyama,
  JHEP {\bf 0405}, 074 (2004)
  [arXiv:hep-th/0312099].

\bibitem{RL} V.~A.~Rubakov,
  arXiv:hep-th/0407104.

\bibitem{D} S.~L.~Dubovsky,
  JHEP {\bf 0410}, 076 (2004)
  [arXiv:hep-th/0409124].

\bibitem{Tin} S.~L.~Dubovsky, P.~G.~Tinyakov and I.~I.~Tkachev,
  Phys.\ Rev.\ Lett.\  {\bf 94}, 181102 (2005)
  [arXiv:hep-th/0411158].

\bibitem{GGrisa} G.~Gabadadze and L.~Grisa,
  Phys.\ Lett.\  B {\bf 617}, 124 (2005)
  [arXiv:hep-th/0412332].


\end{thebibliography}
\end{document}